\documentclass[10pt,twocolumn,oneside,conference]{IEEEtran}

\usepackage{cite}
\usepackage{graphicx}
\usepackage{amsmath}
\usepackage{times}
\usepackage{latexsym}
\usepackage{graphicx}
\usepackage{bm}
\usepackage{amssymb}
\usepackage[center]{caption2}
\usepackage{stfloats}
\usepackage{cases}
\usepackage{array}
\usepackage{setspace}
\usepackage{fancyhdr}
\usepackage{algorithm}
\usepackage{algpseudocode}

\usepackage{amsmath,amsthm}

\usepackage{graphicx}
\usepackage{subfigure}

\usepackage{xcolor}

\usepackage{balance} 
\usepackage{flushend} 

\allowdisplaybreaks[4]

\newcaptionstyle{mystyle1}{%
  \centering TABLE  \captiontext \par}
\captionstyle{mystyle1}
\newcaptionstyle{mystyle2}{%
  \captionlabel $.$ \: \captiontext \par}
\captionstyle{mystyle2}
\newcaptionstyle{mystyle3}{%
  \captionlabel $.$ \: \captiontext \par}
\captionstyle{mystyle3}

\begin{document}

\title{Energy Efficient Power Allocation \\in Massive MIMO Systems based on Standard Interference Function}

\author{
\IEEEauthorblockN{Jiadian Zhang$^\dag$, Yanxiang Jiang$^\dag$$^{*}$, Peng Li$^\dag$, Fuchun Zheng$^\ddag$, and Xiaohu You$^\dag$}
\IEEEauthorblockA{$^\dag$National Mobile Communications Research Laboratory,
Southeast University, Nanjing 210096, China.\\
$^\ddag$School of Systems Engineering, University of
Reading, Reading, RG6 6AY, UK.\\
$^*$E-mail: yxjiang@seu.edu.cn
}
}

\maketitle

\begin{abstract}

        In this paper, energy efficient power allocation for downlink massive MIMO systems is investigated. A constrained non-convex optimization problem is formulated to maximize the energy efficiency (EE), which takes into account the quality of service (QoS) requirements. By exploiting the properties of fractional programming and the lower bound of the user data rate, the non-convex optimization problem is transformed into a convex optimization problem. The Lagrangian dual function method is utilized to convert the constrained convex problem into an unconstrained convex one. Due to the multi-variable coupling problem caused by the intra-user interference, it is intractable to derive an explicit solution to the above optimization problem. Exploiting the standard interference function, we propose an implicit iterative algorithm to solve the unconstrained convex optimization problem and obtain the optimal power allocation scheme. Simulation results show that the proposed iterative algorithm converges in just a few iterations, and demonstrate the impact of the number of users and the number of antennas on the EE.

\end{abstract}

%
%

\section{Introduction}

    With rapid increase in the requirement of data intensive services, huge traffic has been introduced into wireless communication networks in recent years. As a promising candidate, large scale multiple-input multiple-output (also called massive MIMO) technology is proposed to enhance system capacity \cite{Marzetta}. However, the energy consumption in massive MIMO systems is nearly proportional to the number of antennas. Excessive energy consumption of wireless communication networks induces both the increasing carbon emission and unaffordable operational expenditure. As a result, energy efficient system designs have recently drawn much attention.

    An energy efficiency (EE) model involving both the uplink and downlink of a single-cell massive MIMO system was established in \cite{Bjornson}, and the interaction among system parameters was analyzed by means of the convex theory. In \cite{Ng}, joint resource allocation including power allocation, data rate adaptation, antenna allocation, and subcarrier allocation was investigated for an orthogonal frequency division multiple access (OFDMA) downlink massive MIMO network. In \cite{Hien}, the relationship between spectral efficiency (SE) and EE was studied. The convex optimization theory was used to derive the optimal EE with respect to a given SE in \cite{Long}. Within the above literature, the intra-user interference, which couples variables with each other, makes resource allocation a challenging work in the massive MIMO system.

    Motivated by the aforementioned observations, energy efficient power allocation scheme for downlink massive MIMO systems is investigated. Taking the sum user transmit power and user data rate constraints into consideration, we first formulate a constrained non-convex optimization problem. Then, by exploiting the properties of fractional programming and the bound of the user rate, a convex optimization problem is derived. Lagrangian dual function is introduced to transform the constrained problem into a non-constrained one. Based on standard interference function (SIF-based), a low complexity algorithm based on implicit iterations is proposed.

    The rest of the paper is organized as follows. The system model of massive MIMO systems is described in Section II. In Section III, the optimization problem is formulated and the energy efficient power allocation scheme is proposed. In section IV, simulation results are shown. Final conclusions are drawn in Section V.

\section{System model}

    We consider the downlink massive MIMO systems as shown in Fig. 1. It is assumed that there are one base station (BS) with \(M\) antennas and \(K\) single-antenna users {sharing the same resource block (RB). Here, one RB refers to one time-frequency resource block which contains 1 time slot and 12 subcarriers (as in Long Term Evolution (LTE)).} 
\begin{figure}[!h]
\centerline{\includegraphics[width=0.5\textwidth,height=0.2\textheight]{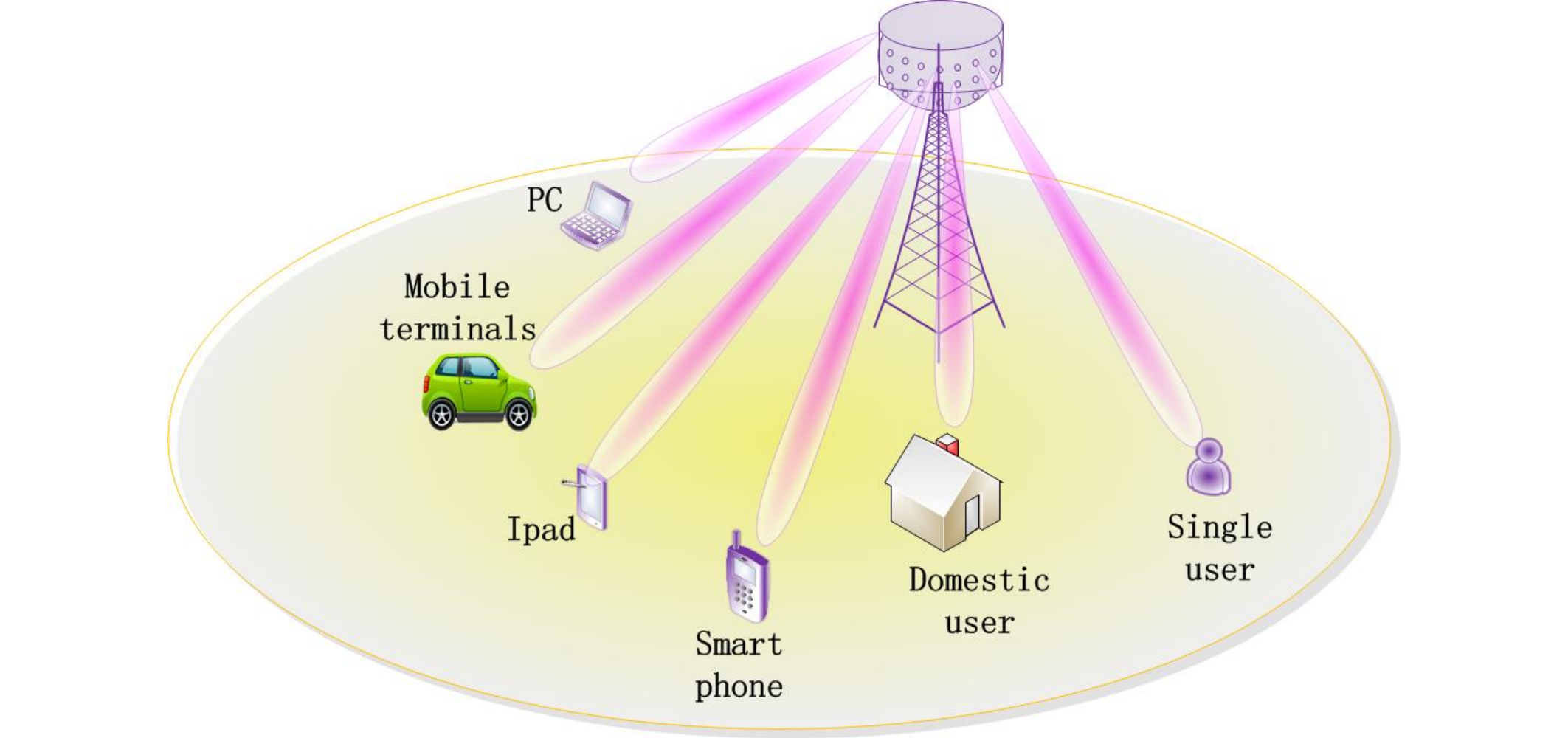}}
\caption{System model of the massive MIMO systems.}
\label{fig1}
\end{figure}

    Let \(\boldsymbol{G}\) denote the flat-fading channel matrix between the BS and the \(K\) users , then it can be written as:
\begin{equation} \label{1}
{\boldsymbol{G}} = {\boldsymbol{H}}{\boldsymbol{D}^{1/2}}
\end{equation}
    where \(\boldsymbol{H}\in {C^{M \times K}}\) is the fast channel matrix with its elements modeled as i.i.d. complex Gaussian random variables, i.e. \(\mathcal{CN}(0,1)\), and \(\boldsymbol{D} = {\rm{diag}}\left\{ {{\beta _1},{\beta _2},\cdots, \beta _k, \cdots,{\beta _K}} \right\}\) denotes the large scale fading matrix with its elements \({\beta _k} = \varphi \varsigma /d_k^\alpha \) which are composed of path loss and shadow fading,  \(\varphi \) is a constant related to the carrier frequency and antenna gain, \({d_k}\) is the distance between the BS and the \(k\textrm{-th}\) user, \(\alpha \) is the path loss exponent, and \(\varsigma \) represents the shadow fading with lognormal distribution \(10{\log _{10}}\varsigma  \sim \mathcal{N}(0,{\sigma ^2})\).

    Let \({y_k}\) denote the received symbol of the \(k\textrm{-th}\) user, then it can be expressed as:
\begin{multline} \label{2}
{y_k} =   \sqrt {{p_k}} \boldsymbol{g}_k^H{\boldsymbol{v}_k}{s_k} + \sum\limits_{\kappa = 1,\kappa \ne k}^K {\sqrt {{p_\kappa}} \boldsymbol{g}_k^H{\boldsymbol{v}_\kappa}{s_\kappa}}  + {n_k}, \\
k \in \left\{ {1,2,...,K} \right\}
\end{multline}
where \({{p_k}}\) is the transmit power allocated to the \(k\textrm{-th}\) user, \({\boldsymbol{g}_k}\) is the \(k\textrm{-th}\) vector of \(\boldsymbol{G}\),
\({\boldsymbol{v}_k}\) is the precoding vector for the \(k\textrm{-th}\) user, \({s_k}\) is the transmit data symbol of the \(k\textrm{-th}\) user, and \({n_k}\) is the additive white gaussian noise (AWGN) at the \(k\textrm{-th}\) user with distribution \(\mathcal{N}(0,{N_0})\), and \({N_0}\) denotes the noise power spectral density.

In order to balance the performance and the complexity, maximum ratio transmitting (MRT) precoding is adopted. Hence, the precoding vector for the \(k\textrm{-th}\) user can be written as \cite{Ng}:
\begin{equation} \label{3}
{\boldsymbol{v}_k} = \frac{{{\boldsymbol{g}_k}}}{{\left\| {{\boldsymbol{g}_k}} \right\|}}
\end{equation}
    where \(\left\|  \cdot  \right\|\) represents the \(L2\)-norm.

    {Let \(B\) denote the bandwidth of one RB. From (\ref{2}) and (\ref{3}), the received signal to interference and noise ratio (SINR) of the \(k\textrm{-th}\) user for a given \({\boldsymbol{g}_k}\) can be expressed as:}
\begin{equation} \label{4}
{\gamma _k} = \frac{{{p_k}{{\left| {{\boldsymbol{g}}_k^H{{\boldsymbol{v}}_k}} \right|}^2}}}{{\sum\limits_{\kappa  = 1,\kappa  \ne k}^K {{p_\kappa }{{\left| {{\boldsymbol{g}}_k^H{{\boldsymbol{v}}_\kappa }} \right|}^2}}  + {BN_0}}}
\end{equation}

    Then, the achievable rate for the \(k\textrm{-th}\) user can be written as:
\begin{equation}\label{4_1}
{r_k} = {{B}\log _2}(1 + {{\gamma _k}})
\end{equation}

    The EE metric is defined as the total average number of bit/Joule successfully delivered to the users. Therefore, the system EE of downlink massive MIMO systems is given by:
\begin{equation} \label{8}
\textrm{EE} = \frac{{\sum\limits_{k = 1}^K {{r_k}} }}{{{\sum\limits_{k = 1}^K {{p_k}}  + \sum\limits_{m = 1}^M {{P_{c,m}}}}}}
\end{equation}
    {where \({P_{c,m}}\) is the constant circuit power consumption per
    antenna. It includes the power dissipations in the baseband processing, transmit
    filter, mixer, frequency synthesizer, and digital-to-analog converter, which are independent of the actual transmit power.}

\section{The proposed power allocation algorithm based on standard interference function}

    In this section, the energy efficient power allocation problem is formulated, and a series of transformation are carried out to obtain a computationally efficient algorithm.

\subsection{Problem formulation}

    The optimal transmit power for all the users can be derived by solving the following optimization problem:
\begin{equation}\begin{split}\label{10}
&\mathop {{\rm{max}}}\limits_{\left\{ {{p_1},{p_2},...,{p_K}} \right\}}{\rm{EE}}\\
\text{s.t.    }
            & C1:\sum\limits_{k = 1}^K {{p_k} \le {P_T}}\\
            & C2:{r_k} \ge {R_{T,k}}, \ k = 1,2...,K\\
\end{split}\end{equation}
    where \({P_{T}}\) denotes the maximum sum transmit power for the BS and \({R_{T,k}}\) denotes the minimum data rate requirement for the \(k\textrm{-th}\) user.

    Note that the optimization  problem in (\ref{10}) is a constrained non-convex optimization problem. {In general, we need an exhaustive search algorithm to obtain the global optimal solution. However, the exhaustive search algorithm has an exponential complexity with respect to (w.r.t.) the number of users, and it is computationally impracticable even for a small size system. Furthermore, tens of users sharing the same RB in massive MIMO systems makes the number of variables in (\ref{10}) increase accordingly.
    Therefore, we propose a series of  transformations  to obtain a computationally efficient power allocation scheme.}

\subsection{Problem transformations}

\subsubsection{Convex transformation}

    The fractional objective function in (\ref{10}) can be classified as a non-linear fractional programming. Let \({q^*}\) denote the maximum EE. By following a similar approach as in \cite{Dinkelbach}, the maximum EE \({q^*}\) can be achieved if and only if:
\begin{equation} \label{11}
\mathop {\max }\limits_{\left\{ {{p_1},{p_2},...,{p_K}} \right\}} \left\{ {\sum\limits_{k = 1}^K {{r_k}}  - {q^*}\left( {\sum\limits_{k = 1}^K {{p_k} + \sum\limits_{m = 1}^M {{P_{c,m}}} } } \right)} \right\} = 0.
\end{equation}

    For massive MIMO systems, the random matrix theory takes effect, as the dimension of channel vector goes high \cite{Tulino}. With perfect channel state information (CSI), Rayleigh fading and MRT precoding, the downlink data rate for the \(k\textrm{-th}\) user can be lower bounded by exploiting the properties of high-dimensional channel vector as follows\cite{Long_2}£º
\begin{equation} \label{6}
{{r_k}\ge{{\hat r}_k} = {{{B}\log }_2}\left( {1 + \frac{{M{\beta _k}{p_k}}}{{{\beta _k}\sum\limits_{\kappa = 1,\kappa \ne k}^K {{p_\kappa}}  + {BN_0}}}} \right)}.
\end{equation}
    {In high SINR region, the received SINR in (\ref{6}) is much larger than 1, therefore, the lower bound of \({r_k}\) can be deduced as:}
\begin{equation}\label{7_1}
{{r_k}\ge{\hat r}_k} \ge {{\tilde r}_k} = {{B}\log _2}\left( {\frac{{M{\beta _k}{p_k}}}{{{\beta _k}\sum\limits_{\kappa = 1,\kappa \ne k}^K {{p_\kappa}}  + {BN_0}}}} \right).
\end{equation}
    {In low SINR region, the lower bound of \({r_k}\) can be deduced as \cite{Venturino}:
    \begin{multline}\label{7_2}
{r_k} \ge {{\hat r}_k} \ge \\
{{\bar r}_k}{\rm{ = }}B\left[ {a{\rm{ + }}b{{\log }_2}\left( {\frac{{M{\beta _k}{p_k}}}{{{\beta _k}\sum\limits_{\kappa  = 1,\kappa  \ne k}^K {{p_\kappa }}  + B{N_0}}}} \right)} \right]
\end{multline}
 where \(a\) and \(b\) are the approximation constants relating to the received SINR.}
 {
 In this paper, we consider the hot spot scenario where high demand on transit data rate is required. Correspondingly, the lower bound in (\ref{7_1}) is adopted in the rest of this paper.
 By comparing (\ref{7_2}) with (\ref{7_1}), it can be observed that the former keeps consistent with the latter in terms of convexity.
 Therefore, energy efficient power allocation scheme in low SINR region can be developed in a similar way by adopting the lower bound in (\ref{7_2}).}

    Let \(q\) denote the value of the EE. By using the lower bound in (\ref{7_1}), the original optimization problem in (\ref{10}) can be simplified as:
\begin{equation}\begin{split}\label{10_10}
\mathop {\max }\limits_{\left\{ {{p_1},{p_2},...,{p_K}} \right\}} &\sum\limits_{k = 1}^K \left\{ {{{\tilde r}_k}}  - q\left( {\sum\limits_{k = 1}^K {{p_k} + \sum\limits_{m = 1}^M {{P_{c,m}}} } } \right) \right\}\\
\text{s.t. \ }
&C1:\sum\limits_{k = 1}^K {{p_k} \le {P_T}} \\
&C2:{{\tilde r}_k} \ge {R_{T,k}}, \ k = 1,2...,K.\\
\end{split}\end{equation}

    The simplified optimization problem in (\ref{10_10}) is a convex problem, which is proved in Appendix A.

\subsubsection{Elimination of constraints}

    The Lagrangian dual function can be utilized to transform the constrained problem in (\ref{10_10}) into an equivalent unconstrained problem{\cite{Ng}}. Let \(\Phi\) be the Lagrangian dual function of (\ref{10_10}), then it can be written as:
\begin{multline}\label{18}
\Phi ({\cal P},\omega ,\boldsymbol{\rho})  = - \left[ {\sum\limits_{k = 1}^K {{{\tilde r}_k}}  - q\left( {\sum\limits_{k = 1}^K {{p_k} + \sum\limits_{m = 1}^M {P_{c,m}} } } \right)} \right]\\
 - \omega \left( {{P_T} - \sum\limits_{k = 1}^K {{p_k}} } \right) - \sum\limits_{k = 1}^K {{\rho_k}\left( {{{\tilde r}_k} - {R_{T,k}}} \right)} \\
 =  - \left[ {\sum\limits_{k = 1}^K {(1 + {\rho_k}){{\tilde r}_k}}  - q\left( {\sum\limits_{k = 1}^K {{p_k} + \sum\limits_{m = 1}^M {P_{c,m}} } } \right)} \right]\\
- w\left( {{P_T} - \sum\limits_{k = 1}^K {{p_k}} } \right) + \sum\limits_{k = 1}^K {{\rho_k}{R_{T,k}}}
\end{multline}
    where \({\cal P}\) denotes the feasible set of power variables, \(\omega \ge 0\) is the Lagrangian multiplier corresponding to the transmit power constraint, and \(\boldsymbol{\rho}\) is the Lagrangian multiplier vector corresponding to the data rate constraints with its element \(\rho_k \ge 0\).

\subsection{Iterative algorithm}

    The Lagrangian dual function in (\ref{18}) can be proved to be convex by exploiting the similar approach in  Appendix A. Therefore, the necessary and sufficient condition to obtain the optimal transmit power can be expressed as:

 \begin{multline}\label{22}
\frac{{\partial \Phi }}{{\partial {p_k}}}= \\ \sum\limits_{\scriptstyle \kappa = 1 \hfill \atop \scriptstyle \kappa \ne k \hfill }^K {\frac{{1 + {\rho_\kappa}}}{{\left( {\sum\limits_{\scriptstyle k' = 1  \hfill \atop \scriptstyle  k' \ne \kappa \hfill}^K {{p_k} + {BN_0}/{\beta _{k'}}} } \right)\ln 2}}}
 - \frac{{1 + {\rho_k}}}{{{p_k}\ln 2}}+ q + \omega
 =0.
 \end{multline}

    From (\ref{22}), it can be seen that it is a challenging work to derive the optimal transmit power in an explicit manner due to the multi-variable coupling problem caused by the intra-user interference.
    Alternatively, the optimal transmit power for the \(k\textrm{-th}\) user can be obtained in an implicit manner as:
\begin{multline} \label{23}
{p_k} = \\
 \frac{{1 + {\rho_k}}}{{\left( {\sum\limits_{\kappa = 1,\kappa \ne k}^K {\frac{{1 + {\rho_\kappa}}}{{\left( {\sum\limits_{k' = 1,k' \ne \kappa}^K {{p_{k'}} + {BN_0}/{\beta _{k'}}} } \right)\ln 2}}}  + q + \omega} \right)\ln 2}}.
\end{multline}
Let the right side hand of (\ref{23}) be denoted by \(T({p_k})\). Then, \(T({p_k})\) can be proved to be a standard interference function with properties of positivity, scalability, and monotonicity,  which is shown in Appendix B.
Therefore, we propose to obtain the optimal transmit power implicitly based on standard interference function.

    Let \(\tau\) denote the predetermined positive threshold for the terminating condition, \(n\) denote the iteration number, \({\theta _1}\) and \({\theta _{2,k}}\) denote the positive step sizes. Then, the SIF-based algorithm is summarized in Algorithm 1. The iteration \(p_k^{(n + 1)} = T(p_k^{(n)})\) is guaranteed to converge to the optimal point quickly \cite{Tan}.

    \begin{algorithm}[!t]
  \caption{The SIF-based Iterative Algorithm}
  \begin{algorithmic}[1]
    \item Initialize the transmit power and Lagrangian multipliers \({{\cal P}^{(0)}},{\omega ^{(0)}},{\boldsymbol{\rho}^{(0)}}\).
    \item Calculate the initial EE \({q^{(0)}} = \frac{{\sum\limits_{k = 1}^K {\tilde r_k^{(0)}} }}{{\sum\limits_{k = 1}^K {p_k^{(0)} + \sum\limits_{m = 1}^M {{P_{c,m}}} } }}\).

        \While {$\sum\limits_{k = 1}^K {\tilde r_k^{(n)}}  - {q^{(n)}}\left( {\sum\limits_{k = 1}^K {p_k^{(n)} + \sum\limits_{m = 1}^M {{P_{c,m}}} } } \right) > \tau  $}
            \For{$i=1:K$}
\begin{multline*}
\ln 2 \cdot p_k^{(n + 1)} = \\
\frac{{1 + \rho _k^{(n)}}}{{ {\sum\limits_{\kappa  = 1,\kappa  \ne k}^K {\frac{{1 + \rho _\kappa ^{(n)}}}{{\left( {\sum\limits_{k' = 1,k' \ne \kappa }^K {p_{k'}^{(n)} + B{N_0}/{\beta _{k'}}} } \right)\ln 2}}}  + {q^{(n)}} + {\omega ^{(n)}}} }}
\end{multline*}

            \EndFor

        Update

        $p_k^{(n)} = p_k^{(n + 1)}$,

        $ {\omega ^{(n + 1)}} = \textrm{max}\left( {0,{\omega ^{(n)}} - {\theta _1}\left( {{P_T} - \sum\limits_{k = 1}^K {p_k^{(n)}} } \right)} \right)$,

        $\rho_k^{(n + 1)} = \textrm{max}\left( {0,\rho_k^{(n)} - {\theta _{2,k}}\left( {\tilde r_k^{(n)} - {R_{T,k}}} \right)} \right)$,

        ${q^{(n+1)}} = \frac{{\sum\limits_{k = 1}^K {\tilde r_k^{(n)}} }}{{\sum\limits_{k = 1}^K {p_k^{(n)} + \sum\limits_{m = 1}^M {{P_{c,m}}} } }}$,

        $n = n + 1$.

    \EndWhile
    \item End
  \end{algorithmic}
\end{algorithm}

    \section{Simulation results}

In this section, the performance of the proposed SIF-based power allocation scheme is evaluated through simulation. A single cell with a radius 500m is considered. The users are uniformly located
within the cell. Without loss of generality, let  \({P_{c,m}} = {P_c}\left( {m = 1,2,...,M} \right)\) and \({R_{T,k}} = {R_T}\left( {k = 1,2,...,K} \right)\). The main parameters used in the simulations are listed in Table I.
\begin{table}[!t]
\centering
\caption{SIMULATION PARAMETERS}
\begin{tabular}{|c|c|}
\hline
Parameter & Value\\
\hline
The radius of the cell & 500m\\
\hline
RB bandwidth $B$ & 120kHz\\
\hline
Number of transmit antennas $M$  & 128\\
\hline
Number of users $K$  & 3\\
\hline
Variance of log-normal shadow fading $\sigma ^2$& 10dB\\
\hline
Factor $\varphi$& 1\\
\hline
Path-loss exponent $\alpha $ & 3.8\\
\hline
Noise spectral density $N_0$ & -170dBm/Hz\\
\hline
Constant power per antenna ${P_{c}}$&0.01W\\
\hline
User rate constraint ${R_{T}}$&6bit/s/Hz\\
\hline
\end{tabular}
\end{table}

In Fig. 2, we show the performance of the SIF-based algorithm compared with that of the two existing algorithms in \cite{Long_2, Runhua}, which are based on convex theory and  game theory, respectively.
 From the figure, it can be observed that the SIF-based algorithm has better performance than that in \cite{Long_2, Runhua}. The reason is that the instantaneous intra-user interference is taken into consideration,
 and the maximum EE can be achieved by the SIF-based algorithm.

\begin{figure}[!t]
\centerline{\includegraphics[width=0.525\textwidth,height=0.296\textheight]{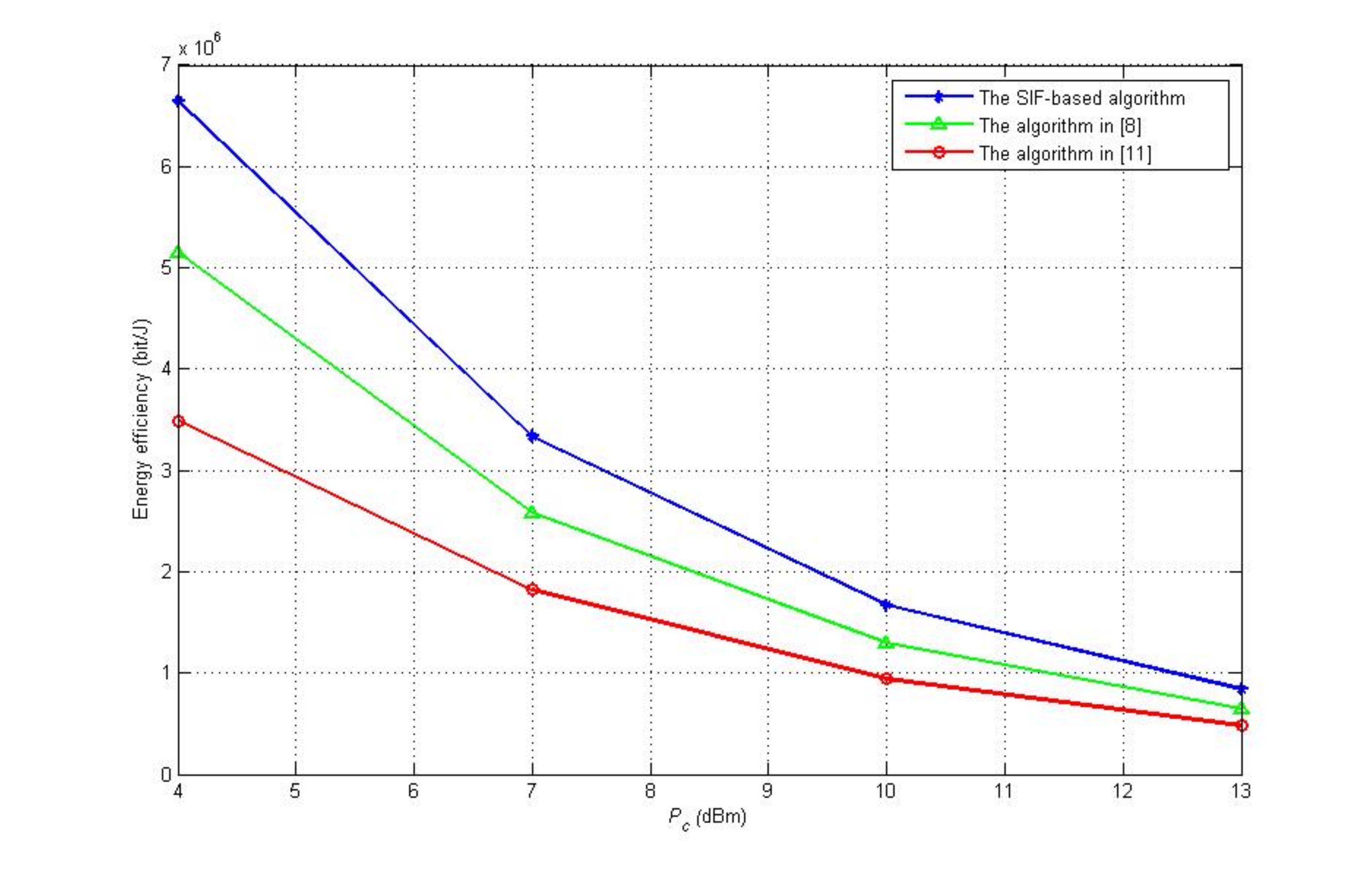}}
\centering
\caption{EE versus different \(P_c\).}
\label{fig1}
\end{figure}

    In Fig. 3, we show the convergency of the SIF-based algorithm and the impact of the transmit power constraint on the system EE.
    It can be seen that the iterative algorithm can always converge to the maximum EE achieved by using the exhaustive search algorithm. {It can also be seen that the larger \(P_T\), the larger EE.
     It can be observed from the red curve in the figure that the SIF-based algorithm intends to achieve a higher EE than the exhaustive search algorithm during the iteration process.
     However, the transmit power is restrained  to be located in the feasible solution. Therefore, the EE value obtained by the SIF-based algorithm finally is lower than that obtained by the exhaustive search algorithm}.

\begin{figure}[!t]
\centerline{\includegraphics[width=0.525\textwidth,height=0.296\textheight]{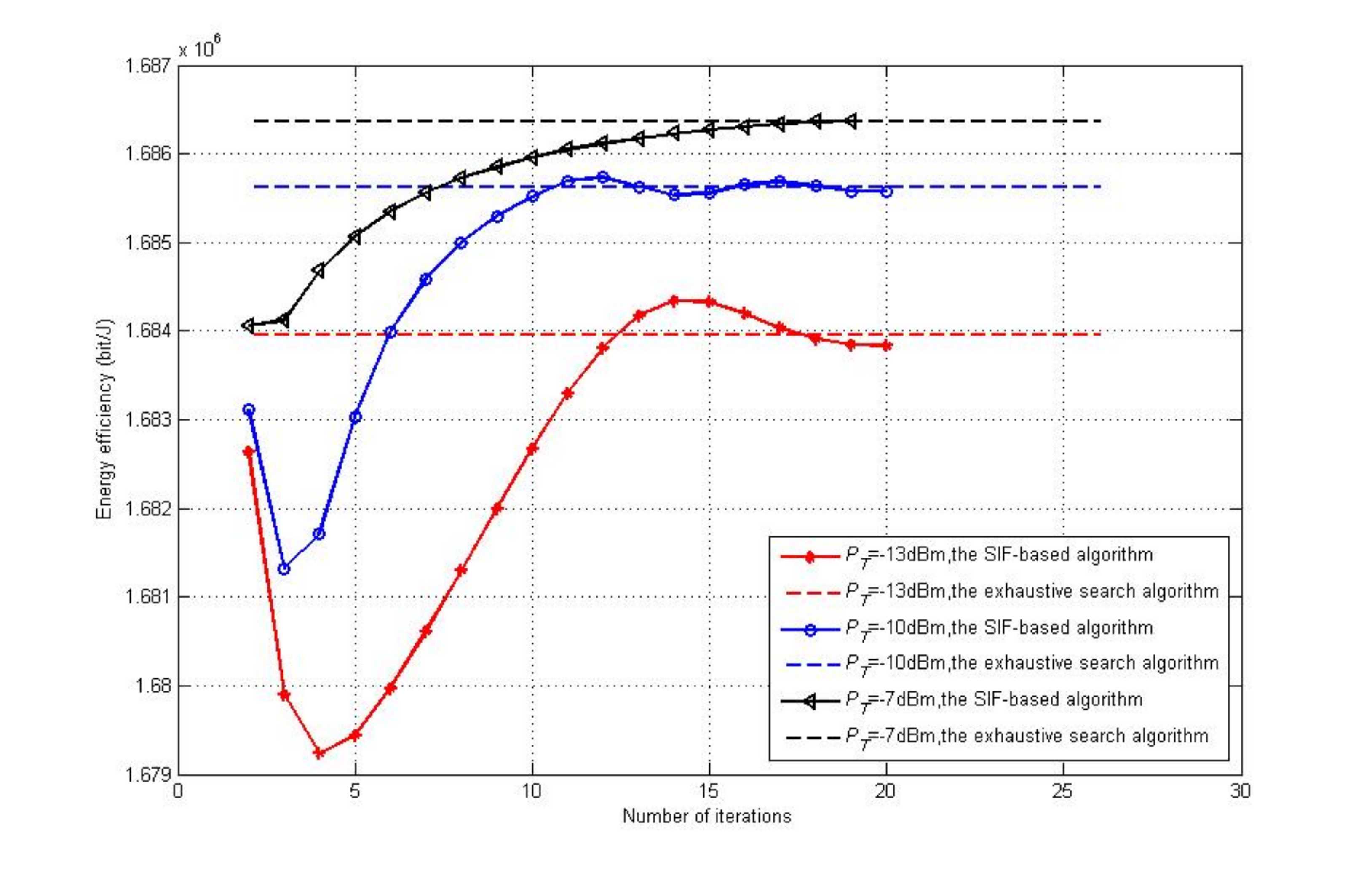}}
\centering
\caption{EE versus the number of iterations with different \(P_T\).}
\label{fig1}
\end{figure}

%
%

    {In Fig. 4, we show how the number of users and that of antennas influence the EE with maximum per-user transmit power 20dBm. Serving more users can always obtain the gain in SE, however, it may be different when referring to EE. We can observe from the figure that there is an optimal number of users leading to the maximum EE for a specific \(M\). We can also observe that more antennas supported more users in order to achieve the maximum EE. Therefore, in future cellular network which contains BSs with hundreds of antennas and a large number of users, it is necessary to combine antenna selection with user management to obtain higher EE.}
    \begin{figure}[!t]
\centerline{\includegraphics[width=0.52\textwidth,height=0.296\textheight]{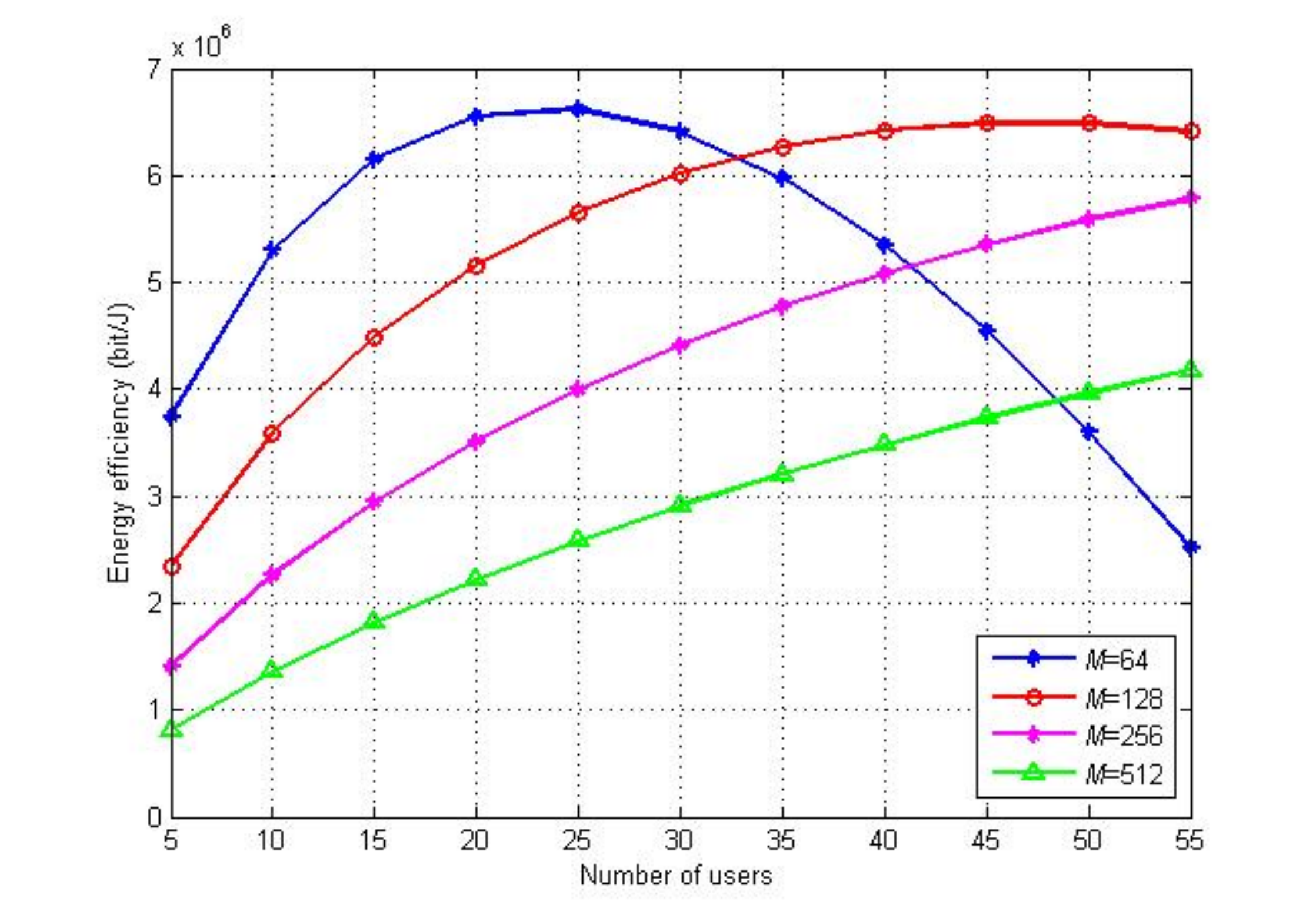}}
\caption{EE versus the number of users under different \(M\) with \(P_c=10\textrm{dBm}\). }
\label{fig1}
\end{figure}

    In Fig. 5, we illustrate how the number of users influence the iterations of the SIF-based algorithm.
    {It can be observed that the number of iterations is nearly linear with the number of users. It indicates that the SIF-based algorithm would guarantee computation efficiency
    even with  a large number of users, compared with the exhaustive search algorithm which has an exponential complexity w.r.t. the number of users.}
\begin{figure}[!t]
\centerline{\includegraphics[width=0.55\textwidth,height=0.291\textheight]{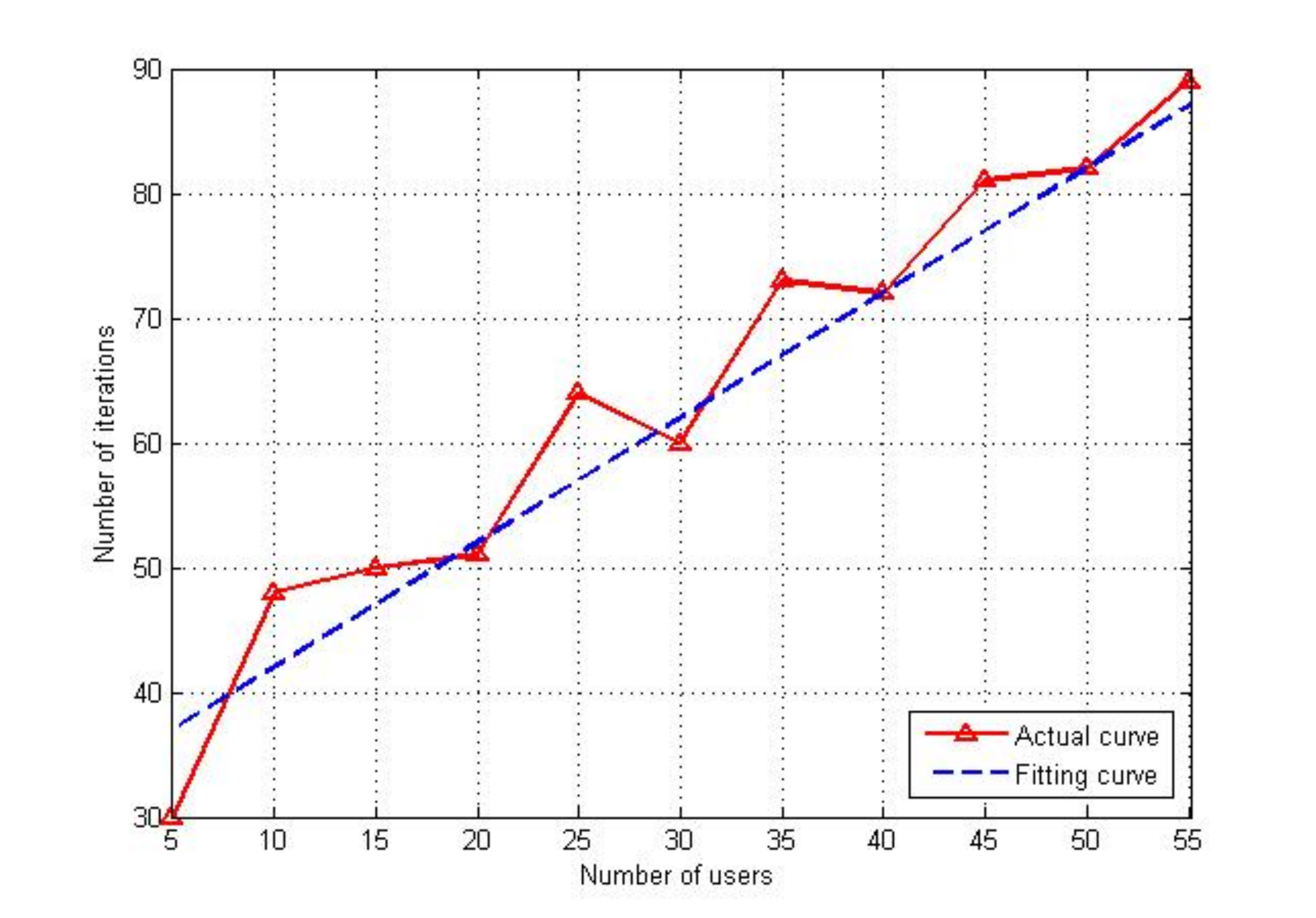}}
\caption{the number of iterations versus the number of users with \(P_c=10\textrm{dBm}\).}
\label{fig1}
\end{figure}


\section{Conclusions}

 In this paper, we have formulated the energy efficient power allocation for massive MIMO systems as a non-convex optimization problem, in which the sum user transmit power and minimum data rate constraints were taken into consideration. A novel iterative method based on standard interference function has been derived. In order to avoid the computational infeasibility arising from intra-user interference, the optimal transmit power has been obtained based on the implicit iteration. Simulation results have shown that the SIF-based algorithm can converge within just a few iterations, and demonstrated the impact of the number of users and the number of antennas on the EE.


\section*{Acknowledgments}
This work was supported in part by the National Basic Research Program of China (973 Program 2012CB316004),
the National 863 Project (2015AA01A709),
and the Natural Science Foundation of China (61221002).
F. Zheng's work is also supported in part by the UK Engineering and Physical Sciences Research Council (EPSRC) under Grant EP/K040685/1.

\section*{Appendix}
\subsection{Proof of the convexity of the problem in (\ref{10_10})}
From  the subtractive EE function in (\ref{11}) , it can be seen the \({{q}\left( {\sum\limits_{k = 1}^K {{p_k} + \sum\limits_{m = 1}^M {{p_{c,m}}} } } \right)}\) is an affine function. As for the \(\sum\limits_{k = 1}^K {{{{\tilde r}_k}}} \) term, by replacing
\({p_k}\) by \({e^{{{\tilde p}_k}}}\) in (\ref{7_1}), it can be derived that:
\begin{multline}
{\sum\limits_{k = 1}^K {\tilde r} _k}{\rm{ }}= \sum\limits_{k = 1}^K {{{{B}\log }_2}\left( {\frac{{M{\beta _k}{e^{{{\tilde p}_k}}}}}{{\sum\limits_{\kappa = 1 \hfill\atop \kappa \ne k}^K {{\beta _k}{e^{{{\tilde p}_\kappa}}} + {N_0}} }}} \right)}  \\
= \sum\limits_{k = 1}^K {{B}\left[ {{{\log }_2}\left( {M{\beta _k}{e^{{{\tilde p}_k}}}} \right) - {{\log }_2}\left( {\sum\limits_{\kappa = 1 \hfill\atop \kappa \ne k}^K {{\beta _k}{e^{{{\tilde p}_\kappa}}} + {BN_0}} } \right)} \right]}
\end{multline}
If we exclude the irrelevant variables in \(\sum\limits_{k = 1}^K {{{\tilde r}_k}} \), the expression corresponding to \({\tilde p_k}\) can be expressed as:
\begin{multline}
f\left( {{{\tilde p}_k}} \right) = \\
{\log _2}\left( {M{\beta _k}{e^{{{\tilde p}_k}}}} \right) - \sum\limits_{\kappa = 1 \hfill\atop \kappa \ne k}^K {{{\log }_2}\left( {\sum\limits_{k' = 1 \hfill\atop k' \ne \kappa}^K {{\beta _{k'}}{e^{{{\tilde p}_{k'}}}} + {BN_0}} } \right)}
\end{multline}
By referring to convex theory \cite{Boyd}, the first term is affine function and the second term is log-sum-exp function, hence \(f({\tilde p_k})\) is concave. Restore the problem by manipulating \({{\tilde p}_k} = \ln \left( {{p_k}} \right)\). Note that log function is concave and the corresponding operation preserves convexity \cite{Boyd}. Therefore,  we can  conclude that the sum rate function \(\sum\limits_{k = 1}^K {{{{\tilde r}_k}}} \) is concave. Consequently, the subtractive EE function in (\ref{10_10}) is concave.
Besides, it can be readily established that C1 is a liner constraint and C2 is a convex constraint.
Therefore, we can draw the conclusion that the simplified optimization problem in  (\ref{10_10}) is a convex problem.

\subsection{Proof of standard interference function in (\ref{23})}
The properties including positivity, monotony and scalability are proved as follows.\\
\emph{{Positivity}}: \(T(p_k)\) can be written as:
\[T\left( {{p_k}} \right) = \frac{{1 + {\rho_k}}}{{\ln 2\left( {\sum\limits_{\kappa = 1 \hfill\atop \kappa \ne k}^K {\frac{{1 + {\rho_{k'}}}}{{\left( {\sum\limits_{k' = 1 \hfill\atop k' \ne \kappa}^K {{p_{k'}} + {BN_0}/{\beta _{k'}}} } \right)\ln 2}}}  + q + \omega} \right)}}\]
It can be seen that both the numerator and denominator keep positive, which guarantee the positivity of \(T(p_k)\).\\
\emph{{Monotony}}: The first-order deviation of \(T(p_k)\) can be obtained as follows:
\begin{multline} \label{24}
\frac{{\partial (T({p_k}))}}{{\partial ({p_k})}} =  \\
- \frac{{1 + {\rho_k}}}{{\ln 2{{\left( {\sum\limits_{\kappa = 1 \hfill\atop \kappa \ne k}^K {\frac{{1 + {\rho_{k'}}}}{{\left( {\sum\limits_{k' = 1 \hfill\atop k' \ne \kappa}^K {{p_{k'}} + {BN_0}/{\beta _{k'}}} } \right)\ln 2}}}  + q + \omega} \right)}^2}}}
\end{multline}
Obviously, the value of the right hand side of (\ref{24}) is less than zero. In turn, \(T(p_k)\) is monotony.\\
\emph{{Scalability}}: \(T(p_k)\) is monotone decreasing. Given a  random number \(\beta  > 1\), it can be proved that:
\[T(\beta {p_k}) < T({p_k}) < \beta T({p_k})\]
Therefore, the scalability is proved.\\

Consequently, \(T(p_k)\) is a standard interference function.
\bibliographystyle{IEEEtran}

\end{document}